\documentclass[letter]{ieice}
\usepackage[dvips]{graphicx}
\usepackage[fleqn]{amsmath}
\usepackage {amssymb,amsthm,amsbsy}
\usepackage[varg]{txfonts}
\usepackage{enumerate}
\field{A}
\title{On the Nonexistence of the Ding-Helleseth-Martinsen' s Constructions of Almost Difference Set for Cyclotomic Classes of Order 6}
\authorlist{% fill arguments of \authorentry, otherwise error will be caused. 
\authorentry[mlqiecully@163.com]{Minglong Qi}{m}{etab1}
\authorentry{Shengwu Xiong}{n}{etab1}
\authorentry{Jingling Yuan}{n}{etab1}
\authorentry{Wenbi Rao}{n}{etab1}
\authorentry{Luo Zhong}{n}{etab1}
}
\affiliate[etab1]{School of Computer Science and Technology, Wuhan University of Technology, Mafangshan West Campus, 430070 Wuhan City, China }

\newtheorem{sec3_thm1}{Theorem}%[section]

\newtheorem{sec3_lemma1}{Lemma} %[section]
\newtheorem{sec3_lemma2}[sec3_lemma1]{Lemma}
\newtheorem{sec3_lemma3}[sec3_lemma1]{Lemma}
\newtheorem{sec3_lemma4}[sec3_lemma1]{Lemma}
\newtheorem{sec3_lemma5}[sec3_lemma1]{Lemma}
\newtheorem{sec3_lemma6}[sec3_lemma1]{Lemma}
\newtheorem{sec3_lemma7}[sec3_lemma1]{Lemma}
\newtheorem{sec3_lemma8}[sec3_lemma1]{Lemma}
\newtheorem{sec3_lemma9}[sec3_lemma1]{Lemma}
\newtheorem{sec3_lemma10}[sec3_lemma1]{Lemma}
\newtheorem{sec3_lemma11}[sec3_lemma1]{Lemma}
\newtheorem{sec3_lemma12}[sec3_lemma1]{Lemma}
\begin{document}
\maketitle
\begin{summary}
Pseudorandom sequences with optimal three-level autocorrelation have important applications in CDMA communication systems. Constructing the sequences with three-level autocorrelation is equivalent to finding cyclic almost difference sets as their supports. In a paper of Ding, Helleseth, and Martinsen, the authors developed a new method known as the Ding-Helleseth-Martinsen’s Constructions in literature to construct the almost difference set using product set between $ GF(2) $ and union sets of cyclotomic classes of order 4. In this correspondence, we show that there do not exist such constructions for cyclotomic classes of order 6.
\end{summary}

\section{Introduction}\label{sec 1}
Let  $ (A,+) $ be an Abelian group with $ n $ elements and $ D $ be a $ k$-subset of $ A $. Define the distance function $ d_{D}(e)=|(D+e)\cap D| $, where $ D+e=\lbrace x+e\,|\,x\in D \ \text{and}\  e\in D\setminus \lbrace 0\rbrace\rbrace$. $ D $ is referred to as an $ (n,\,k,\,\lambda,\,t) $ almost difference set if $ d_{D}(e) $ takes on the value $ \lambda $ altogether $ t $ times and on the value $ \lambda+1 $ altogether $ n-1-t $ times when $ e $  ranges over all the nonzero elements of $ A $. Let $ q=df+1 $ be a power of an odd prime, $ \alpha $ be a primitive element of extension field $ \mathit{GF}(q) $. Define the cosets $ \mathit{D}^{(d,q)}_{i}=\lbrace\alpha^{kd+i}\,|\,0\leq k<f \rbrace,\ 0\leq i<d $, which are called the cyclotomic classes of order $ d $ with respect to $ \mathit{GF}(q) $. It is obvious that $\mathit{GF}(q)^{*}=\bigcup_{i=0}^{d-1} \mathit{D}^{(d,q)}_{i} $. The constants $ (m,\;n)_{d}=|(\mathit{D}^{(d,q)}_{m}+1)\cap \mathit{D}^{(d,q)}_{n}| $ are known as the cyclotomic numbers of order $ d $ with respect to $ \mathit{GF}(q) $.

Pseudorandom sequences with cyclic almost difference sets as their support sets find important applications in CDMA systems \cite{ar05,ar06}. In \cite{ar03,ar06}, the authors developed a new method known as the Ding-Helleseth-Martinsen’ s Constructions in literature to construct the almost difference set using product sets between $ GF(2) $ and union sets from the cyclotomic classes of order 4. In this letter, we show that there do not exist such constructions for cyclotomic classes of order 6. The rest of the letter is structured as follows: in Section \ref{sec 2}, the cyclotomic numbers of order 6 and their corresponding formulae are presented; in Section \ref{sec 3}, the main theorem of the present letter is given and proved; finally a brief concluding remark is given in Section \ref{sec 4}.
\section{Cyclotomic Numbers of Order Six }\label{sec 2}
%\subsection{Cyclotomic Numbers of Order Six}\label{subsec-2.1}
Let $ p=6f+1 $ be an odd prime with $ f $ even. It is well known that $ p $ can be expanded to $ p=A^{2}+3B^{2} $. Even though there are 36 cyclotomic numbers of order 6, but there are only ten irreducible ones which can be expressed in linear combination of the vector $ <p,A,B,1>^{T} $ \cite{ar01,ar02}. The relations of the 36 cyclotomic numbers of order 6 with respect to the ten irreducible ones are listed in Table \ref{lab-table-cyclotomic-number-order6}. From Table \ref{lab-table-cyclotomic-number-order6}, it is easy to see that, for instance, $ (2,5)_{6} =(1,3)_{6}$. Given the prime $ p $, and its decomposed parameters $ A $ and $ B $, the ten distinct cyclotomic numbers of order 6 of $ p $ can be calculated by the formulae exhibited in Table \ref{lab-table-formulae-for-cn-order6}, but there being three different sets of the formulae determined by the residue of $ m $ modulo 3, where $ \alpha^{m}\equiv 2 \pmod p $ with $ \alpha $ a primitive root of $ p $.
\begin{table}[tb]
\caption{The relations of cyclotomic numbers of order 6}
\label{lab-table-cyclotomic-number-order6}
{\renewcommand{\tabcolsep}{0.15cm}
\begin{center}
\begin{tabular}{|c|c|c|c|c|c|c|}
\hline
(h,k) & 0 & 1 & 2 & 3 & 4 & 5 \\
\hline
0 & (0, 0) & (0, 1) & (0, 2) & (0, 3) & (0, 4) & (0, 5) \\
\hline
1 & (0, 1) & (0, 5) & (1, 2) & (1, 3) & (1, 4) & (1, 2) \\
\hline
2 & (0, 2) & (1, 2) & (0, 4) & (1, 4) & (2, 4) & (1, 3) \\
\hline
3 & (0, 3) & (1, 3) & (1, 4) & (0, 3) & (1, 3) & (1, 4) \\
\hline
4 & (0, 4) & (1, 4) & (2, 4) & (1, 3) & (0, 2) & (1, 2) \\
\hline
5 & (0, 5) & (1, 2) & (1, 3) & (1, 4) & (1, 2) & (0, 1) \\
\hline
\end{tabular}
\end{center}
}
\end{table}

\begin{table}[tb]
\caption{The cyclotomic numbers of order 6 for $ f $ even}
\label{lab-table-formulae-for-cn-order6}
{\renewcommand{\tabcolsep}{0.15cm}
\begin{center}
\begin{tabular}{|c|c|c|c|}
\hline
& $ m\equiv 0\pmod 3 $    &   $ m\equiv 1\pmod 3 $ & $ m\equiv 2\pmod 3 $\\
\hline
36(0, 0)&	p-17-20A	&	p-17-8A+6B	    &	p-17-8A-6B\\
\hline
36(0, 1)&	p-5+4A+18B	&	p-5+4A+12B	    &	p-5+4A+6B\\
\hline
36(0, 2)&	p-5+4A+6B	&	p-5+4A-6B	    &	p-5-8A\\
\hline
36(0, 3)&	p-5+4A  	&	p-5+4A-6B 	    &	p-5+4A+6B\\
\hline
36(0, 4)&	p-5+4A-6B	&	p-5-8A   	    &	p-5+4A+6B\\
\hline
36(0, 5)&	p-5+4A-18B	&	p-5+4A-6B	    &	p-5+4A-12B\\
\hline
36(1, 2)&	p+1-2A  	&	p+1-2A-6B	    &	p+1-2A+6B\\
\hline
36(1, 3)&	p+1-2A  	&	p+1-2A-6B	    &	p+1-2A-12B\\
\hline
36(1, 4)&	p+1-2A  	&	p+1-2A+12B	    &	p+1-2A+6B\\
\hline
36(2, 4)&	p+1-2A  	&	p+1+10A+6B	    &	p+1+10A-6B\\
\hline
\end{tabular}
\end{center}
}
\end{table}
\section{Nonexistence of the DHM Constructions for Cyclotomic Classes of Order 6}\label{sec 3}
Let $ \mathbb{S}_{n}^{k} $ denote the set of all the $ k $-subsets of $ Z_{n} $ with $ 1\leq k< n $. 
% \begin{sec2_def1}\label{ads-definition}
% 
%  \end{sec2_def1} 
Throughout the rest of the present letter, the following notation is kept unchanged. Let $ p=6f+1 $ be  an odd prime with $ f $ even, $ D_{i}^{(6,p)} $ denote the $ i^{th} $ cyclotomic class of order 6 with $ 0\leq i<6 $, $ I,J\subset \mathbf{Z}_{6} $ be index subsets. Define $ \mathit{D}_{I}=\bigcup_{i\in I}D_{i}^{(6,p)},\  \mathit{D}_{J}=\bigcup_{j\in J}D_{j}^{(6,p)},\ \mathit{C}=\lbrace 0\rbrace\times  \mathit{D}_{I}\cup \lbrace 1\rbrace\times  \mathit{D}_{J}$ and $ ,\mathit{C}^{'}=\lbrace 0\rbrace\times  \mathit{D}_{I}\cup \lbrace 1\rbrace\times  \mathit{D}_{J}\cup \lbrace(0,\,0) \rbrace$. Define also the following distance functions
 \begin{equation*}
 \begin{split}
 d_{I}(w)&=|( \mathit{D}_{I}+w)\cap  \mathit{D}_{I}|,\\
 d_{I,J}(w)&=|( \mathit{D}_{I}+w)\cap  \mathit{D}_{J}|,\\
 d_{\mathit{C}}(w_{1},\,w_{2})&=|\bigl(\mathit{C}+(w_{1},\,w_{2}) \bigr)\cap \mathit{C}|,\\
 d_{\mathit{C}^{'}}(w_{1},\,w_{2})&=|\bigl(\mathit{C}^{'}+(w_{1},\,w_{2}) \bigr)\cap \mathit{C}^{'}|,
 \end{split}
 \end{equation*}
 where $ w,\ w_{2} \in Z_{p}$ and $ w_{1} \in \mathit{GF}(2) $. It is clear that  $ p $ can be expressed as $ p=A^{2}+3B^{2}$\cite{ar01,ar02}. 
%  Let 
%   \begin{equation*}
%   \begin{split}
%   \mathit{Q}&= D_{0}^{(6,q)}\cup  D_{2}^{(6,q)}\cup  D_{4}^{(6,q)},\\
%   \mathit{QN}&= D_{1}^{(6,q)}\cup  D_{3}^{(6,q)}\cup  D_{5}^{(6,q)}.
%   \end{split}
%   \end{equation*}
  
 The distance functions $  d_{\mathit{C}}(w_{1},\,w_{2}) $ and $ d_{\mathit{C}^{'}}(w_{1},\,w_{2}) $ can be explicitly expanded out in $ d_{I}(w_{2})  $, $ d_{J}(w_{2})  $ and $ d_{I,J}(w_{2}) $, stated by the following two lemmas whose proofs can be found in \cite[eq.(2) and eq.(4)]{ar03}
  \begin{sec3_lemma1}\label{lab-sec3-lamma1}
   \begin{equation*}
   d_{\mathit{C}}(w_{1},\,w_{2})=\\
   \begin{cases}
   |\mathit{D}_{I}|+|\mathit{D}_{J}|&\quad\mbox{if}\ w_{1}=0,w_{2}=0\\
   d_{I}(w_{2})+d_{J}(w_{2})&\quad\mbox{if}\ w_{1}=0,w_{2}\ne 0\\
   d_{I,J}(w_{2})+d_{J,I}(w_{2})&\quad\mbox{if}\ w_{1}=1,w_{2}\ne 0\\
   2|\mathit{D}_{I}\cap \mathit{D}_{J}|&\quad\mbox{if}\ w_{1}=1,w_{2}= 0
   \end{cases}
   \end{equation*}
   \end{sec3_lemma1}
 \begin{sec3_lemma2}\label{lab-sec3-lamma2}
  \begin{equation*}
  \begin{split}
   &d_{\mathit{C}^{'}}(w_{1},\,w_{2})=d_{\mathit{C}}(w_{1},\,w_{2})\\
   &+
    \begin{cases}
      |\mathit{D}_{I}\cap \lbrace w_{2},\,-w_{2}\rbrace|&\quad\mbox{if}\ w_{1}=0,w_{2}\ne 0\\
      |\mathit{D}_{J}\cap \lbrace w_{2},\,-w_{2}\rbrace|&\quad\mbox{if}\ w_{1}=1,w_{2}\ne 0\\
     0 &\quad\mbox{otherwise}.
    \end{cases}
  \end{split}
  \end{equation*}
  \end{sec3_lemma2}
  
  For the following lemmas and theorems of this section,  let $ w\in Z_{p}^{*} $ and $ w^{-1} \in  D_{h}^{(6,p)}$, denote $ d_{I}(w) $ by  $ d_{I}(h) $ where $ I\in Z_{6}$ is an index subset.
%  it is always supposed that $ p=12f+1 $ being an odd prime and is expanded to $ A^{2}+3B^{2} $, $ \alpha $ is a primitive root of $ p $, and $ \alpha^{m}\equiv 2\pmod p $. Especially, the meaning of the notation $ m $ is fixed. In addition, let $ w\in Z_{p}^{*} $ and $ w^{-1} \in  D_{h}^{(6,p)}$, denote $ d_{I}(w) $ by  $ d_{I}(h) $ where $ I\in Z_{6}$ is an index subset.

\begin{sec3_lemma3}\label{sec3-lamma3-label}
Let $ I=\lbrace0,1,2\rbrace \in  \mathbb{S}_{6}^{3}$. Then, the distance function $ d_{I}(w) $ can be calculated using the following formulae:
\begin{itemize}
\item Case $ m\equiv 0 \pmod 3 $.
\begin{equation*}
d_{I}(w)=\\                  
\begin{cases}
\frac{p}{4}+\frac{2B}{3}-\frac{5}{4}&\quad \mbox{if}\ w^{-1} \in  D_{0}^{(6,p)},\\
\frac{p}{4}-\frac{2B}{3}-\frac{1}{4}&\quad \mbox{if}\ w^{-1} \in  D_{1}^{(6,p)},\\
\frac{p}{4}-\frac{1}{4}&\quad \mbox{if}\ w^{-1} \in  D_{2}^{(6,p)},\\
\frac{p}{4}+\frac{2B}{3}-\frac{1}{4}&\quad \mbox{if}\ w^{-1} \in  D_{3}^{(6,p)},\\
\frac{p}{4}-\frac{2B}{3}-\frac{5}{4}&\quad \mbox{if}\ w^{-1} \in  D_{4}^{(6,p)},\\
\frac{p}{4}-\frac{5}{4}&\quad \mbox{if}\ w^{-1} \in  D_{5}^{(6,p)}.
\end{cases}
\end{equation*}
\item Case $ m\equiv 1 \pmod 3 $.
\begin{equation*}
d_{I}(w)=\\                  
\begin{cases}
\frac{p}{4}-\frac{5}{4}&\quad \mbox{if}\ w^{-1} \in  D_{0}^{(6,p)},\\
\frac{p}{4}-\frac{A+B}{3}-\frac{1}{4}&\quad \mbox{if}\ w^{-1} \in  D_{1}^{(6,p)},\\
\frac{p}{4}+\frac{A+B}{3}-\frac{1}{4}&\quad \mbox{if}\ w^{-1} \in  D_{2}^{(6,p)},\\
\frac{p}{4}-\frac{1}{4}&\quad \mbox{if}\ w^{-1} \in  D_{3}^{(6,p)},\\
\frac{p}{4}-\frac{A+B}{3}-\frac{5}{4}&\quad \mbox{if}\ w^{-1} \in  D_{4}^{(6,p)},\\
\frac{p}{4}-\frac{A+B}{3}+\frac{5}{4}&\quad \mbox{if}\ w^{-1} \in  D_{5}^{(6,p)}.
\end{cases}
\end{equation*}
\item Case $ m\equiv 2 \pmod 3 $.
\begin{equation*}
d_{I}(w)=\\                  
\begin{cases}
\frac{p}{4}-\frac{A-B}{3}-\frac{5}{4}&\quad \mbox{if}\ w^{-1} \in  D_{0}^{(6,p)},\\
\frac{p}{4}-\frac{1}{4}&\quad \mbox{if}\ w^{-1} \in  D_{1}^{(6,p)},\\
\frac{p}{4}+\frac{A-B}{3}-\frac{1}{4}&\quad \mbox{if}\ w^{-1} \in  D_{2}^{(6,p)},\\
\frac{p}{4}-\frac{A-B}{3}-\frac{1}{4}&\quad \mbox{if}\ w^{-1} \in  D_{3}^{(6,p)},\\
\frac{p}{4}-\frac{5}{4}&\quad \mbox{if}\ w^{-1} \in  D_{4}^{(6,p)},\\
\frac{p}{4}+\frac{A-B}{3}-\frac{5}{4}&\quad \mbox{if}\ w^{-1} \in  D_{5}^{(6,p)}.
\end{cases}
\end{equation*}
\end{itemize}
\end{sec3_lemma3}               
 \begin{proof}
 We only prove the case $ m\equiv 0 \pmod 3 $.
 \begin{equation}\label{sec3-lemma3-proof}
 \begin{split}
  d_{I}(w)&=|( \mathit{D}_{I}+w)\cap  \mathit{D}_{I}|\\
   &=|(\bigcup_{i\in I}D_{i}^{(6,p)}+w)\cap (\bigcup_{j\in I}D_{j}^{(6,p)})|\\
   &=|\bigl(\bigcup_{i\in I}(D_{i}^{(6,p)}+w)\bigr)\cap (\bigcup_{j\in I}D_{j}^{(6,p)})|\\
   &=|\bigl(\bigcup_{i\in I}(w^{-1}D_{i}^{(6,p)}+1)\bigr)\cap \bigcup_{j\in I}w^{-1}D_{j}^{(6,p)}|\\
    &=|\bigl(\bigcup_{i\in I}(D_{i+h}^{(6,p)}+1)\bigr)\cap (\bigcup_{j\in I}D_{j+h}^{(6,p)})|\\
  &=\sum_{i\in I}\sum_{j\in I}|(D_{i+h}^{(6,p)}+1)\cap D_{j+h}^{(6,p)}|\\
  &=\sum_{i\in I}\sum_{j\in I}(i+h,\,j+h)_{6}\\
  &=(h,\,h)_{6}+(h,\,1+h)_{6}+(h,\,2+h)_{6}+(1+h,\,h)_{6}+\\
  &\ \quad (1+h,\,1+h)_{6}+(1+h,\,2+h)_{6}+(2+h,\,h)_{6}+\\
  &\ \quad (2+h,\,1+h)_{6}+(2+h,\,2+h)_{6}.
  \end{split}
 \end{equation}
Making varying $ h $ from 0 to 5 in the last equation of eq.(\ref{sec3-lemma3-proof}), we can obtain the following formula. Recall that all the involved subscripts should be reduced modulo 6.
\begin{equation}\label{sec3-lemma3-proof02}
 d_{I}(h)=\\
 \begin{cases}
&(0,0)+(0,1)+(0,2)+(1,0)+(1,1)+\\
&(1,2)+(2,0)+(2,1)+(2,2)\quad \mbox{if}\ h=0;\\
&(1,1)+(1,2)+(1,3)+(2,1)+(2,2)+\\
&(2,3)+(3,1)+(3,2)+(3,3)\quad \mbox{if}\ h=1;\\
&(2,2)+(2,3)+(2,4)+(3,2)+(3,3)+\\
&(3,4)+(4,2)+(4,3)+(4,4)\quad \mbox{if}\ h=2;\\
&(3,3)+(3,4)+(3,5)+(4,3)+(4,4)+\\
&(4,5)+(5,3)+(5,4)+(5,5)\quad \mbox{if}\ h=3;\\
&(4,4)+(4,5)+(4,0)+(5,4)+(5,5)+\\
&(5,0)+(0,4)+(0,5)+(0,0)\quad \mbox{if}\ h=4;\\
&(5,5)+(5,0)+(5,1)+(0,5)+(0,0)+\\
&(0,1)+(1,5)+(1,0)+(1,1)\quad \mbox{if}\ h=5.
 \end{cases}
\end{equation}
Remark that in eq.(\ref{sec3-lemma3-proof02}) the subscript 6 is omitted for all the cyclotomic numbers due to limited displaying. Using Table \ref{lab-table-cyclotomic-number-order6} to reduce all the cyclotomic numbers occurring in eq.(\ref{sec3-lemma3-proof02}) into the ten irreducible ones leads to a new form to eq.(\ref{sec3-lemma3-proof02}):
\begin{equation}\label{sec3-lemma3-proof03}
 d_{I}(h)=\\
 \begin{cases}
 &(0,0)+2(0,1)+2(0,2)+(0,4)+(0,5)+\\
 & 2(1,2) \quad \mbox{if}\ h=0;\\
 &(0,3)+(0,4)+(0,5)+2(1,2)+2(1,3)+\\
 & 2(1,4) \quad \mbox{if}\ h=1;\\
 &(0,2)+(0,3)+(0,4)+2(1,3)+2(1,4)+\\
 & 2(2,4) \quad \mbox{if}\ h=2;\\
 &(0,1)+(0,2)+(0,3)+2(1,2)+2(1,3)+\\
 & 2(1,4) \quad \mbox{if}\ h=3;\\
 &(0,0)+(0,1)+(0,2)+2(0,4)+2(0,5)+\\
 & 2(1,2) \quad \mbox{if}\ h=4;\\
 &(0,0)+3(0,1)+3(0,5)+2(1,2)\\ 
 &\quad \mbox{if}\ h=5.
 \end{cases}
\end{equation}
Last step of the proof consists of substituting the ten formulae in the first column of Table \ref{lab-table-formulae-for-cn-order6} for the corresponding cyclotomic numbers occurring in eq.(\ref{sec3-lemma3-proof03}) after which the assertion of Lemma \ref{sec3-lamma3-label} follows.
 \end{proof} 
 
 In order to compute the distance function $  d_{\mathit{C}}(w_{1},\,w_{2}) $ (See the beginning of Section \ref{sec 3}) the following several lemmas are directly written down of which the proof is quite similar to that for Lemma \ref{sec3-lamma3-label} and omitted.
 
\begin{sec3_lemma4}\label{sec3-lamma4-label}
Let $ J=\lbrace0,4,5\rbrace \in  \mathbb{S}_{6}^{3}$. Then, the distance function $ d_{J}(w) $ can be calculated using the following formulae:
\begin{itemize}
\item Case $ m\equiv 0 \pmod 3 $.
\begin{equation*}
d_{J}(w)=\\                  
\begin{cases}
\frac{p}{4}-\frac{2B}{3}-\frac{5}{4}&\quad \mbox{if}\ w^{-1} \in  D_{0}^{(6,p)},\\
\frac{p}{4}-\frac{5}{4}&\quad \mbox{if}\ w^{-1} \in  D_{1}^{(6,p)},\\
\frac{p}{4}+\frac{2B}{3}-\frac{5}{4}&\quad \mbox{if}\ w^{-1} \in  D_{2}^{(6,p)},\\
\frac{p}{4}-\frac{2B}{3}-\frac{1}{4}&\quad \mbox{if}\ w^{-1} \in  D_{3}^{(6,p)},\\
\frac{p}{4}-\frac{1}{4}&\quad \mbox{if}\ w^{-1} \in  D_{4}^{(6,p)},\\
\frac{p}{4}+\frac{2B}{3}-\frac{1}{4}&\quad \mbox{if}\ w^{-1} \in  D_{5}^{(6,p)}.
\end{cases}
\end{equation*}
\item Case $ m\equiv 1 \pmod 3 $.
\begin{equation*}
d_{J}(w)=\\                  
\begin{cases}
\frac{p}{4}-\frac{A+B}{3}-\frac{5}{4}&\quad \mbox{if}\ w^{-1} \in  D_{0}^{(6,p)},\\
\frac{p}{4}+\frac{A+B}{3}-\frac{5}{4}&\quad \mbox{if}\ w^{-1} \in  D_{1}^{(6,p)},\\
\frac{p}{4}-\frac{5}{4}&\quad \mbox{if}\ w^{-1} \in  D_{2}^{(6,p)},\\
\frac{p}{4}-\frac{A+B}{3}-\frac{1}{4}&\quad \mbox{if}\ w^{-1} \in  D_{3}^{(6,p)},\\
\frac{p}{4}+\frac{A+B}{3}-\frac{1}{4}&\quad \mbox{if}\ w^{-1} \in  D_{4}^{(6,p)},\\
\frac{p}{4}-\frac{1}{4}&\quad \mbox{if}\ w^{-1} \in  D_{5}^{(6,p)}.
\end{cases}
\end{equation*}
\item Case $ m\equiv 2 \pmod 3 $.
\begin{equation*}
d_{J}(w)=\\                  
\begin{cases}
\frac{p}{4}-\frac{5}{4}&\quad \mbox{if}\ w^{-1} \in  D_{0}^{(6,p)},\\
\frac{p}{4}+\frac{A-B}{3}-\frac{5}{4}&\quad \mbox{if}\ w^{-1} \in  D_{1}^{(6,p)},\\
\frac{p}{4}-\frac{A-B}{3}-\frac{5}{4}&\quad \mbox{if}\ w^{-1} \in  D_{2}^{(6,p)},\\
\frac{p}{4}-\frac{1}{4}&\quad \mbox{if}\ w^{-1} \in  D_{3}^{(6,p)},\\
\frac{p}{4}+\frac{A-B}{3}-\frac{1}{4}&\quad \mbox{if}\ w^{-1} \in  D_{4}^{(6,p)},\\
\frac{p}{4}-\frac{A-B}{3}-\frac{1}{4}&\quad \mbox{if}\ w^{-1} \in  D_{5}^{(6,p)}.
\end{cases}
\end{equation*}
\end{itemize}
\end{sec3_lemma4}            

\begin{sec3_lemma5}\label{sec3-lamma5-label}
Let $ I=\lbrace0,1,2\rbrace \in  \mathbb{S}_{6}^{3}$ and $ J=\lbrace0,4,5\rbrace \in  \mathbb{S}_{6}^{3}$. Then, the distance function $ d_{I,J}(w) $ can be calculated using the following formulae:
\begin{itemize}
\item Case $ m\equiv 0 \pmod 3 $.
\begin{equation*}
d_{I,J}(w)=\\                  
\begin{cases}
\frac{p}{4}-\frac{A}{3}-\frac{11}{12}&\quad \mbox{if}\ w^{-1} \in  D_{0}^{(6,p)},\\
\frac{p}{4}+\frac{A+B}{6}-\frac{5}{12}&\quad \mbox{if}\ w^{-1} \in  D_{1}^{(6,p)},\\
\frac{p}{4}+\frac{A-B}{6}-\frac{5}{12}&\quad \mbox{if}\ w^{-1} \in  D_{2}^{(6,p)},\\
\frac{p}{4}-\frac{A}{3}+\frac{1}{12}&\quad \mbox{if}\ w^{-1} \in  D_{3}^{(6,p)},\\
\frac{p}{4}+\frac{A+B}{6}-\frac{5}{12}&\quad \mbox{if}\ w^{-1} \in  D_{4}^{(6,p)},\\
\frac{p}{4}+\frac{A-B}{6}-\frac{5}{12}&\quad \mbox{if}\ w^{-1} \in  D_{5}^{(6,p)}.
\end{cases}                      
\end{equation*}
\item Case $ m\equiv 1 \pmod 3 $.
\begin{equation*}
d_{I,J}(w)=\\                  
\begin{cases}
\frac{p}{4}+\frac{B}{3}-\frac{11}{12}&\quad \mbox{if}\ w^{-1} \in  D_{0}^{(6,p)},\\
\frac{p}{4}+\frac{A-3B}{6}-\frac{5}{12}&\quad \mbox{if}\ w^{-1} \in  D_{1}^{(6,p)},\\
\frac{p}{4}-\frac{A-B}{6}-\frac{5}{12}&\quad \mbox{if}\ w^{-1} \in  D_{2}^{(6,p)},\\
\frac{p}{4}+\frac{B}{3}+\frac{1}{12}&\quad \mbox{if}\ w^{-1} \in  D_{3}^{(6,p)},\\
\frac{p}{4}+\frac{A-3B}{6}-\frac{5}{12}&\quad \mbox{if}\ w^{-1} \in  D_{4}^{(6,p)},\\
\frac{p}{4}-\frac{A-B}{6}-\frac{5}{12}&\quad \mbox{if}\ w^{-1} \in  D_{5}^{(6,p)}.
\end{cases}                               
\end{equation*}
\item Case $ m\equiv 2 \pmod 3 $.
\begin{equation*}
d_{I,J}(w)=\\                  
\begin{cases}
\frac{p}{4}-\frac{B}{3}-\frac{11}{12}&\quad \mbox{if}\ w^{-1} \in  D_{0}^{(6,p)},\\
\frac{p}{4}-\frac{A+B}{6}-\frac{5}{12}&\quad \mbox{if}\ w^{-1} \in  D_{1}^{(6,p)},\\
\frac{p}{4}+\frac{A+3B}{6}-\frac{5}{12}&\quad \mbox{if}\ w^{-1} \in  D_{2}^{(6,p)},\\
\frac{p}{4}-\frac{B}{3}+\frac{1}{12}&\quad \mbox{if}\ w^{-1} \in  D_{3}^{(6,p)},\\
\frac{p}{4}-\frac{A+B}{6}-\frac{5}{12}&\quad \mbox{if}\ w^{-1} \in  D_{4}^{(6,p)},\\
\frac{p}{4}+\frac{A+3B}{6}-\frac{5}{12}&\quad \mbox{if}\ w^{-1} \in  D_{5}^{(6,p)}.
\end{cases}
\end{equation*}
\end{itemize}
\end{sec3_lemma5}
\begin{sec3_lemma6}\label{sec3-lemma6}
Let $ I=\lbrace0,1,2\rbrace \in  \mathbb{S}_{6}^{3}$ and $ J=\lbrace0,4,5\rbrace \in  \mathbb{S}_{6}^{3}$. Then, $ d_{I,J}(w)=d_{J,I}(w) $ for $ w\in Z_{p}^{*} $.
\end{sec3_lemma6}   

Now, we are ready to compute the distance function $ d_{\mathit{C}}(w_{1},\,w_{2}) $.

\begin{sec3_lemma7}\label{sec3-lemma7}
Let $ I=\lbrace0,1,2\rbrace \in  \mathbb{S}_{6}^{3}$ and $ J=\lbrace0,4,5\rbrace \in  \mathbb{S}_{6}^{3}$. Then, the distance function, $ d_{\mathit{C}}(w_{1},\,w_{2})=|\bigl(\mathit{C}+(w_{1},\,w_{2}) \bigr)\cap \mathit{C}| $, can be calculated by the following formulae:
\begin{itemize}
\item Case $ m\equiv 0 \pmod 3 $.
\begin{itemize}
\item $ w_{1}=0 $.
\begin{equation*}
d_{\mathit{C}}(w_{1},\,w_{2})=\\
\begin{cases}
\frac{p-5}{2}&\quad \mbox{if}\  w^{-1}_{2} \in  D_{0}^{(6,p)},\\
\frac{3p-4B-9}{6}&\quad \mbox{if}\  w^{-1}_{2} \in  D_{1}^{(6,p)},\\
\frac{3p+4B-9}{6}&\quad \mbox{if}\  w^{-1}_{2} \in  D_{2}^{(6,p)},\\
\frac{p-1}{2}&\quad \mbox{if}\  w^{-1}_{2} \in  D_{3}^{(6,p)},\\
\frac{3p-4B-9}{6}&\quad \mbox{if}\  w^{-1}_{2} \in  D_{4}^{(6,p)},\\
\frac{3p+4B-9}{6}&\quad \mbox{if}\  w^{-1}_{2} \in  D_{5}^{(6,p)}.
\end{cases}
\end{equation*}
\item $ w_{1}=1 $.
\begin{equation*}
d_{\mathit{C}}(w_{1},\,w_{2})=\\
\begin{cases}
\frac{3p-4A-11}{6}&\quad \mbox{if}\  w^{-1}_{2} \in  D_{0}^{(6,p)},\\
\frac{3p+2A+2B-5}{6}&\quad \mbox{if}\  w^{-1}_{2} \in  D_{1}^{(6,p)},\\
\frac{3p+2A-2B-5}{6}&\quad \mbox{if}\  w^{-1}_{2} \in  D_{2}^{(6,p)},\\
\frac{3p-4A+1}{6}&\quad \mbox{if}\  w^{-1}_{2} \in  D_{3}^{(6,p)},\\
\frac{3p+2A+2B-5}{6}&\quad \mbox{if}\  w^{-1}_{2} \in  D_{4}^{(6,p)},\\
\frac{3p+2A-2B-5}{6}&\quad \mbox{if}\  w^{-1}_{2} \in  D_{5}^{(6,p)}.
\end{cases}
\end{equation*}
\item $ w_{1}=1 $ and $ w_{2}=0 $.
\begin{equation*}
d_{\mathit{C}}(w_{1},\,w_{2})=\frac{p-1}{3}.
\end{equation*}
\end{itemize}                           
\item Case $ m\equiv 1 \pmod 3 $.
\begin{itemize}
\item $ w_{1}=0 $.
\begin{equation*}
d_{\mathit{C}}(w_{1},\,w_{2})=\\
\begin{cases}
\frac{3p-2A-2B-15}{6}&\quad \mbox{if}\  w^{-1}_{2} \in  D_{0}^{(6,p)},\\
\frac{p-3}{2}&\quad \mbox{if}\  w^{-1}_{2} \in  D_{1}^{(6,p)},\\
\frac{3p+2A+2B-9}{6}&\quad \mbox{if}\  w^{-1}_{2} \in  D_{2}^{(6,p)},\\
\frac{3p-2A-2B-3}{6}&\quad \mbox{if}\  w^{-1}_{2} \in  D_{3}^{(6,p)},\\
\frac{p-3}{2}&\quad \mbox{if}\  w^{-1}_{2} \in  D_{4}^{(6,p)},\\
\frac{3p+2A+2B-9}{6}&\quad \mbox{if}\  w^{-1}_{2} \in  D_{5}^{(6,p)}.
\end{cases}
\end{equation*}
\item $ w_{1}=1 $.
\begin{equation*}
d_{\mathit{C}}(w_{1},\,w_{2})=\\
\begin{cases}
\frac{3p+4B-11}{6}&\quad \mbox{if}\  w^{-1}_{2} \in  D_{0}^{(6,p)},\\
\frac{3p+2A-6B-5}{6}&\quad \mbox{if}\  w^{-1}_{2} \in  D_{1}^{(6,p)},\\
\frac{3p-2A+2B-5}{6}&\quad \mbox{if}\  w^{-1}_{2} \in  D_{2}^{(6,p)},\\
\frac{3p+4B+1}{6}&\quad \mbox{if}\  w^{-1}_{2} \in  D_{3}^{(6,p)},\\
\frac{3p+2A-6B-5}{6}&\quad \mbox{if}\  w^{-1}_{2} \in  D_{4}^{(6,p)},\\
\frac{3p-2A+2B-5}{6}&\quad \mbox{if}\  w^{-1}_{2} \in  D_{5}^{(6,p)}.
\end{cases} 
\end{equation*}
\item $ w_{1}=1 $ and $ w_{2}=0 $.
\begin{equation*}
d_{\mathit{C}}(w_{1},\,w_{2})=\frac{p-1}{3}.
\end{equation*}
\end{itemize}                    
\item Case $ m\equiv 2 \pmod 3 $.
\begin{itemize}
\item $ w_{1}=0 $.
\begin{equation*}
d_{\mathit{C}}(w_{1},\,w_{2})=\\
\begin{cases}
\frac{3p-2A+2B-15}{6}&\quad \mbox{if}\  w^{-1}_{2} \in  D_{0}^{(6,p)},\\
\frac{3p+2A-2B-9}{6}&\quad \mbox{if}\  w^{-1}_{2} \in  D_{1}^{(6,p)},\\
\frac{p-3}{2}&\quad \mbox{if}\  w^{-1}_{2} \in  D_{2}^{(6,p)},\\
\frac{3p-2A+2B-3}{6}&\quad \mbox{if}\  w^{-1}_{2} \in  D_{3}^{(6,p)},\\
\frac{3p+2A-2B-9}{6}&\quad \mbox{if}\  w^{-1}_{2} \in  D_{4}^{(6,p)},\\
\frac{p-3}{2}&\quad \mbox{if}\  w^{-1}_{2} \in  D_{5}^{(6,p)}.
\end{cases}
\end{equation*}
\item $ w_{1}=1 $.
\begin{equation*}
d_{\mathit{C}}(w_{1},\,w_{2})=\\
\begin{cases}
\frac{3p-4B-11}{6}&\quad \mbox{if}\  w^{-1}_{2} \in  D_{0}^{(6,p)},\\
\frac{3p-2A-2B-5}{6}&\quad \mbox{if}\  w^{-1}_{2} \in  D_{1}^{(6,p)},\\
\frac{3p+2A+6B-5}{6}&\quad \mbox{if}\  w^{-1}_{2} \in  D_{2}^{(6,p)},\\
\frac{3p-4B+1}{6}&\quad \mbox{if}\  w^{-1}_{2} \in  D_{3}^{(6,p)},\\
\frac{3p-2A-2B-5}{6}&\quad \mbox{if}\  w^{-1}_{2} \in  D_{4}^{(6,p)},\\
\frac{3p+2A+6B-5}{6}&\quad \mbox{if}\  w^{-1}_{2} \in  D_{5}^{(6,p)}.
\end{cases}                   
\end{equation*}
\item $ w_{1}=1 $ and $ w_{2}=0 $.
\begin{equation*}
d_{\mathit{C}}(w_{1},\,w_{2})=\frac{p-1}{3}.
\end{equation*}
\end{itemize}           
\end{itemize}
\end{sec3_lemma7}
\begin{proof}
The actual lemma can be proved by using Lemma \ref{lab-sec3-lamma1} as the leading lemma and Lemmas \ref{lab-sec3-lamma2}-\ref{sec3-lemma6} as auxiliary lemmas.
\end{proof}

\begin{sec3_lemma8}\label{sec3-lemma8}
Let $ I=\lbrace0,1,2\rbrace \in  \mathbb{S}_{6}^{3}$ and $ J=\lbrace0,4,5\rbrace \in  \mathbb{S}_{6}^{3}$. Define $ \mathit{D}_{I}=\bigcup_{i\in I}D_{i}^{(6,p)},\  \mathit{D}_{J}=\bigcup_{j\in J}D_{j}^{(6,p)},\ \mathit{C}=\lbrace 0\rbrace\times  \mathit{D}_{I}\cup \lbrace 1\rbrace\times  \mathit{D}_{J}$. Then, $ \mathit{C} $ cannot form an almost difference set over $ GF(2)\times Z_{p} $.
\end{sec3_lemma8}                                      
\begin{proof}
Lemma \ref{sec3-lemma7} gives the distance function $ d_{\mathit{C}}(w_{1},\,w_{2}) $. Its value distribution according to all the cases of $ w_{2} $ is so irregular that the condition of forming the almost difference set specified in Section \ref{sec 1} cannot be fulfilled for whatever are the parameters $ A $ and $ B $.
\end{proof}
\begin{sec3_lemma9}\label{sec3-lemma9}
Let $ I, J \in  \mathbb{S}_{6}^{3}$,  $ \mathit{D}_{I}=\bigcup_{i\in I}D_{i}^{(6,p)},\  \mathit{D}_{J}=\bigcup_{j\in J}D_{j}^{(6,p)},\ \mbox{and}\ \mathit{C}=\lbrace 0\rbrace\times  \mathit{D}_{I}\cup \lbrace 1\rbrace\times  \mathit{D}_{J}$. Then, $ \mathit{C} $ cannot form an almost difference set over $ GF(2)\times Z_{p} $.
\end{sec3_lemma9}
\begin{proof}
For each pair of subscript sets $ (I,J)\in  \mathbb{S}_{6}^{3}\times   \mathbb{S}_{6}^{3}$, the distance function $ d_{\mathit{C}}(w_{1},\,w_{2}) $ can be computed out as for  Lemma \ref{sec3-lemma9}. Computational results show that there are no pair of subscript sets $ (I,J)\in  \mathbb{S}_{6}^{3}\times   \mathbb{S}_{6}^{3}$ such that  $ \mathit{C} $  forms an almost difference set over $ GF(2)\times Z_{p} $.
\end{proof}
\begin{sec3_lemma10}\label{sec3-lemma10}
Let $ I, J \in  \mathbb{S}_{6}^{k}$ where $ 1\leq k<6 $,  $ \mathit{D}_{I}=\bigcup_{i\in I}D_{i}^{(6,p)},\  \mathit{D}_{J}=\bigcup_{j\in J}D_{j}^{(6,p)},\ \mbox{and}\ \mathit{C}=\lbrace 0\rbrace\times  \mathit{D}_{I}\cup \lbrace 1\rbrace\times  \mathit{D}_{J}$. Then, $ \mathit{C} $ cannot form an almost difference set over $ GF(2)\times Z_{p} $.
\end{sec3_lemma10}
\begin{proof}
Similar to the proof for Lemma \ref{sec3-lemma9}.
\end{proof}

\begin{sec3_lemma11}\label{sec3-lemma11}
Let $ p=12f+1 $ be an odd prime, and $ I \in  \mathbb{S}_{6}^{k}$ with $ 1\leq k<6 $. If $ D_{0}^{(6,p)}\subset \mathit{D}_{I} $ then $ \delta_{I}= |\mathit{D}_{I}\cap \lbrace w_{2},\,-w_{2}\rbrace|=2 $, else  $ \delta_{I}= |\mathit{D}_{I}\cap \lbrace w_{2},\,-w_{2}\rbrace|=0$. Where $ w_{2}\in Z_{p}^{*} $.
\end{sec3_lemma11} 
\begin{proof}
It is obvious.
\end{proof}

\begin{sec3_lemma12}\label{sec3-lemma12}
Let $ I, J \in  \mathbb{S}_{6}^{k}$ where $ 1\leq k<6 $,  $ \mathit{D}_{I}=\bigcup_{i\in I}D_{i}^{(6,p)},\  \mathit{D}_{J}=\bigcup_{j\in J}D_{j}^{(6,p)},\ \mbox{and}\ \mathit{C}^{'}=\lbrace 0\rbrace\times  \mathit{D}_{I}\cup \lbrace 1\rbrace\times  \mathit{D}_{J} \cup \lbrace (0,0)\rbrace$. Then, $ \mathit{C}^{'} $ cannot form an almost difference set over $ GF(2)\times Z_{p} $.
\end{sec3_lemma12}
\begin{proof}
The present lemma can be proved by using Lemma \ref{lab-sec3-lamma2} as the leading lemma and Lemma \ref{sec3-lemma10}, Lemma \ref{sec3-lemma11} as auxiliary lemmas.
\end{proof}

We are now at the step to be able to assert whether or not there exist the DHM Constructions for the cyclotomic classes of order 6.

\begin{sec3_thm1}\label{sec3-theorem1}
Let $ p=12f+1 $ be an odd prime, and suppose that $ p=A^{2}+3B^{2} $. Then, there are no the DHM Constructions of the almost difference set from product sets between $ GF(2) $ and union sets of cyclotomic classes of order 6 for the prime $ p $.
\end{sec3_thm1}
\begin{proof}
By Lemma \ref{sec3-lemma10} and Lemma \ref{sec3-lemma12}.
\end{proof}

\section{Conclusion}\label{sec 4}
Pseudorandom sequences with optimal three-level autocorrelation have important applications in CDMA communication. Constructing such sequences is equivalent to finding cyclic almost difference sets as their supports. The Ding-Helleseth-Martinsen’s Constructions is an efficient method to construct the almost difference set. In this letter it is shown that there are no such constructions for the cyclotomic classes of order 6.
%\bibliographystyle{ieicetr}% bib style
%\bibliography{}% your bib database

\end{document}